\documentclass[runningheads]{llncs}
\usepackage{graphicx}
\usepackage{xcolor}
\usepackage{ltxtable}
\usepackage{multicol}
\usepackage{multirow,url}
\usepackage[misc]{ifsym} 
\begin{document}
\title{A Two-Stage Cascade Model with Variational Autoencoders and Attention Gates for MRI Brain Tumor Segmentation}
%
%
\author{Chenggang~Lyu \and 
Hai~Shu$^{(\mbox{\scriptsize{\Letter}})}$}
\authorrunning{C. Lyu and H. Shu}
%
\institute{Department of Biostatistics, School of Global Public Health, New York University, New York, NY 10003, USA \\
\email{hs120@nyu.edu}}
\maketitle              
\begin{abstract}
Automatic MRI brain tumor segmentation is of vital importance for the disease diagnosis, monitoring, and treatment planning. In this paper, we propose a two-stage encoder-decoder based model for brain tumor subregional segmentation.  
Variational autoencoder regularization is utilized in both stages to prevent the overfitting issue. The second-stage network adopts attention gates and is trained additionally using an expanded dataset formed by  the first-stage outputs. On the BraTS~2020 validation dataset, the proposed method achieves the mean Dice score of 0.9041, 0.8350, and 0.7958, and Hausdorff distance (95\%) of 4.953 , 6.299, 23.608 for the whole tumor, tumor core, and enhancing tumor,{\tiny } respectively. The corresponding results on the BraTS~2020 testing dataset are 0.8729, 0.8357, and 0.8205 for Dice score, and 11.4288, 19.9690, and 15.6711 for Hausdorff distance.
The code is publicly available at
\url{https://github.com/shu-hai/two-stage-VAE-Attention-gate-BraTS2020}.

\end{abstract}
\section{Introduction}
Brain tumors can be categorized into primary tumors and secondary tumors depending on where they originate. Glioma, the most common type of primary brain tumor, can be further categorized into low-grade gliomas (LGG) and high-grade gliomas (HGG). HGG is a malignant brain tumor type with a high degree of aggressiveness that often requires surgery. Usually, several complimentary 3D Magnetic Resonance Imaging (MRI) modalities are acquired to highlight different tissue properties and areas of tumor spread. Compared to traditional methods that rely on physicians’ professional knowledge and experience, automatic 3D brain tumor segmentation is time-efficient and can provide objective and reproducible results for further tumor analysis and monitoring. In recent years,  deep-learning based segmentation approaches have exhibited superior performance than traditional methods.

\begin{figure}[h!]
	\centering
	\includegraphics[width=1\textwidth]{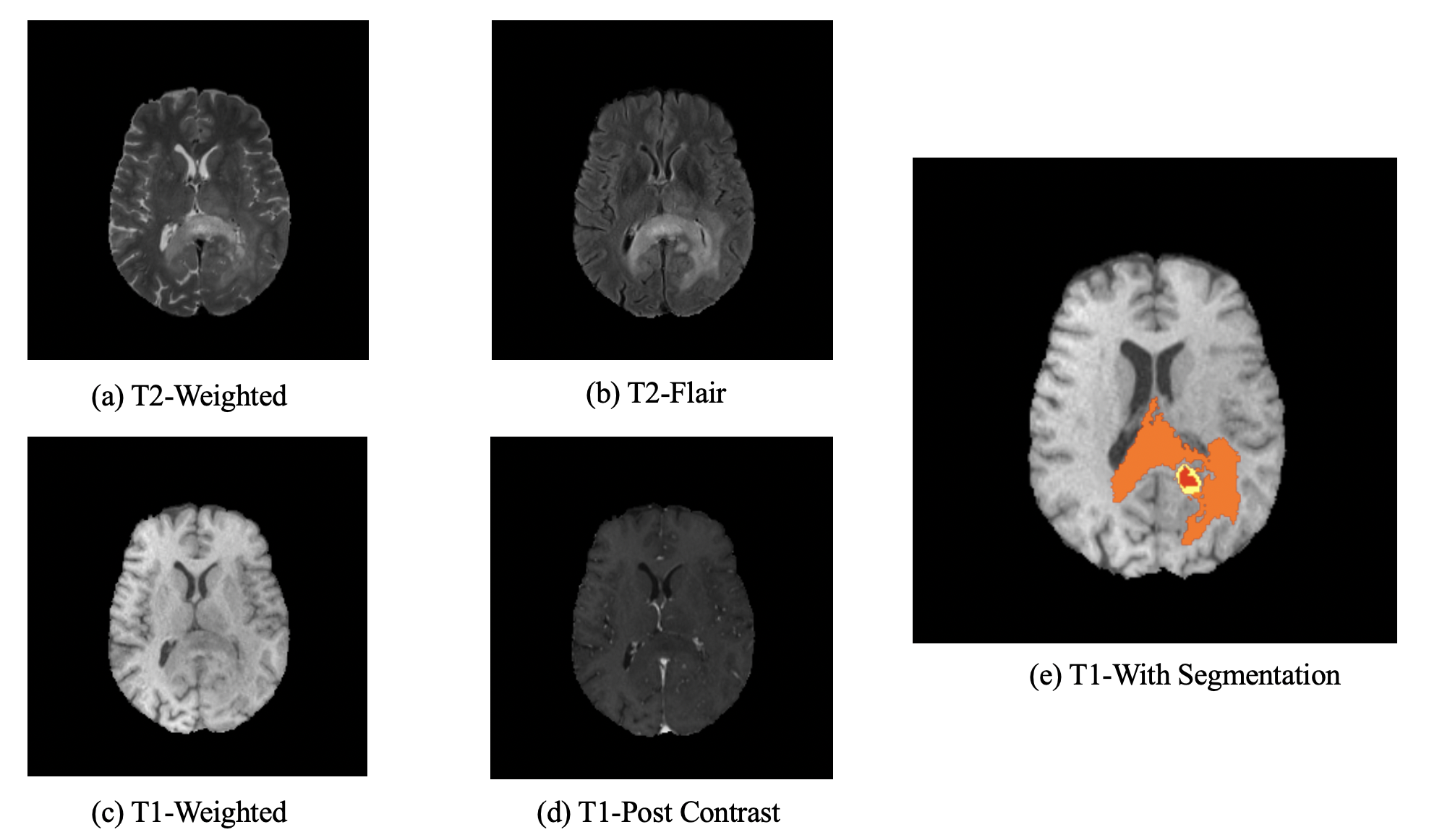}
	\caption{An example image set. Subfigure~(e) highlights three tumor subregions: ED (orange), NCR/NET (yellow), and ET (red).}
	\label{fig: example}
\end{figure}

The Multimodal Brain Tumor Segmentation Challenge (BraTS) is an annual international competition that aims to evaluate state-of-the-art methods of brain tumor segmentation \cite{bakagbm2017,bakalgg2017,bakas2017advancing,menze2015multimodal}. The organizer provides a 3D multimodal MRI
dataset with ``ground-truth" tumor segmentation labels annotated by physicians and radiologists. For each patient, four 3D MRI modalities are provided 
including native T1-weighted (T1), post-contrast T1-weighted (T1c), T2-weighted (T2), and T2 Fluid Attenuated Inversion Recovery (T2-FLAIR) volumes. The
brain tumor
 segmentation task concentrates on three tumor sub-regions: the necrotic and non-enhancing tumor (NCR/NET, labeled 1), the peritumoral edema (ED, labeled 2) and the  GD-enhancing tumor (ET, labeled 4). Fig.~\ref{fig: example} shows an image set of a patient. 
The rankings of competing methods for this segmentation task are determined by metrics, including Dice score, Hausdorff distance (95\%), Sensitivity, and Specificity, evaluated on the testing dataset for ET, tumor core (TC=ET+NCR/NET), and whole tumor (WT=TC+ED) \cite{Spyridon2018summary}.

In BraTS~2018, Myronenko \cite{myronenko20183d} proposed an asymmetrical U-Net with a larger encoder for feature extraction and a smaller decoder for label reconstruction, and won the first place of the challenge. An encouraging innovation of the method is utilizing a variational autoencoder (VAE) branch to regularize the encoder and boost generalization performance.
The champion team of BraTS~2019,
Jiang et al. \cite{jiang2019two}, proposed a two-stage network, which used an asymmetrical U-Net, similar to Myronenko \cite{myronenko20183d}, in the first stage to obtain a coarse prediction, and then 
employed
a similar but wider network in the second stage to refine the prediction. An additional branch was adopted in the decoder of the second-stage network to regularize the associated encoder. The success of the above two models indicates the feasibility and the importance of adding a branch to the decoder to reduce overfitting and boost the model performance. 

Compared with general computer vision problems, 3D MRI image segmentation tasks generally face two special challenges: the scarcity of training data and the class imbalance~\cite{zhao2019bag}. To alleviate the shortage of training data, Isensee et al.~\cite{isensee2018no} took the advantage of additional labeled data by using a co-training strategy. Zhou et al.~\cite{zhou2018one} combined several performance-boosting tricks, such as introducing a focal loss to alleviate the class imbalance, to achieve further improvements. 

For brain tumor segmentation tasks specifically, another challenging difficulty is the variability of tumor morphology and location across different tumor development stages and different cases.  To improve the prediction accuracy, many segmentation methods \cite{wang2017automatic,tu2009auto,zhou2018one,zhou2018learning} decompose the task into separate localization and subsequent segmentation steps, with additional preceding models for object localization.  For instance, Wang et al.~\cite{wang2017automatic}  sequentially trained three networks according to the tumor subregion hierarchy. Oktay et al.~\cite{oktay2018attention} demonstrated that the same objective can be achieved by introducing attention gates (AGs) into the standard convolutional-neural-network framework in pancreas tumor segmentation tasks.

Inspired by aforementioned works, in this paper we propose a two-stage cascade network for brain tumor segmentation. We borrow the network structure of Myronenko~\cite{myronenko20183d} as the first-stage network to obtain relatively rough segmentation results. The second stage network uses the concatenation of the preliminary segmentation maps from the first-stage network and the MRI images as the input, with the aim to refine the prediction of the NCR/NET and ET subregions. 
We apply AGs \cite{oktay2018attention} to further suppress the feature responses in irrelevant background regions. 
Our second-stage network exhibits the capabilities to (i) provide more model candidates with  competitive performance for model ensembling, (ii) stabilize the predictions across models of different epochs, and (iii) improve the performance of each single model, particularly for NCR/NET and ET.
The implementation details and segmentation results are provided in Sections~\ref{Sec: Experiment} and~\ref{Sec: results}.

\section{Method}\label{Sec: method}
The proposed two-stage network structure consists of two cascaded networks.
The first-stage network takes the multimodal MRI images as input and predicts coarse segmentation maps. The concatenation of the preliminary segmentation maps and the MRI images is passed into the second-stage network to generate improved segmentation results. 

\subsection{The First-Stage Network:  Asymmetrical U-Net with a VAE Branch}
\label{sect:Stage1}
The network architecture (Fig.~\ref{fig: Stage1&2}) consists of a larger encoding path for semantic feature extraction, a smaller decoding path for  segmentation map prediction, and a VAE branch for input images reconstruction. This part is identical to the network proposed in \cite{myronenko20183d}.

\subsubsection{Encoder}
The encoder consists of ResNet~\cite{he2016deep,he2016identity} blocks for four spatial levels, with the number of blocks 1, 2, 2, and 4, respectively.  Each ResNet block has two convolutions with Group Normalization and ReLU, followed by an additive identity skip connection. The input of the encoder is an MRI crop of size $4{\times}160{\times}192{\times}128$, with the first channel referring to the four MRI modalities. The input is processed by a $3{\times}3{\times}3$ convolution layer with 32 filters and a dropout layer with a rate of 0.2, and then passed through a series of ResNet blocks. Between every two blocks with different spatial levels, a $3{\times}3{\times}3$ convolution with a stride of 2 is used to reduce the resolution of the feature maps by 2 and double the number of feature channels simultaneously. The endpoint of the encoder has size $256{\times}20{\times}24{\times}16$, which is 1/8 of the spatial size of the input data.

\subsubsection{Decoder}
The decoder has an almost symmetrical architecture with the encoder, except for the number of ResNet blocks within each spatial level is 1. After each block, we use a trilinear up-sampler to recover the spatial size by 2 and a $1 {\times}1{\times}1$ convolution to reduce the number of feature channels by 2, followed by an additive skip connection from the encoder output of the corresponding spatial level. The operations within each block are the same as those in the encoder. At the end of the decoder, a $1{\times}1{\times}1$ convolution is used to reduce the number of feature channels from 32 to 3, followed by a sigmoid function to convert feature maps into probability maps.

\subsubsection{VAE Branch}
This decoder branch receives the output of the encoder and produces a reconstructed image of the original input. In the beginning, the decoder endpoint output is reduced to a lower-dimensional space of 256 using a fully connected layer, where 256 represents 128 means and 128 standard deviations of Gaussian distributions, from which a sample of size 128 is drawn. Then the drawn vector is mapped back to the high-dimensional space with the same spatial property and reconstructed into the input image dimensions gradually following the same strategy as the decoder. Notice that there is no additive skip connection between encoder and the VAE branch.

\begin{figure}[b!]
	\centering
	\includegraphics[width=1\textwidth]{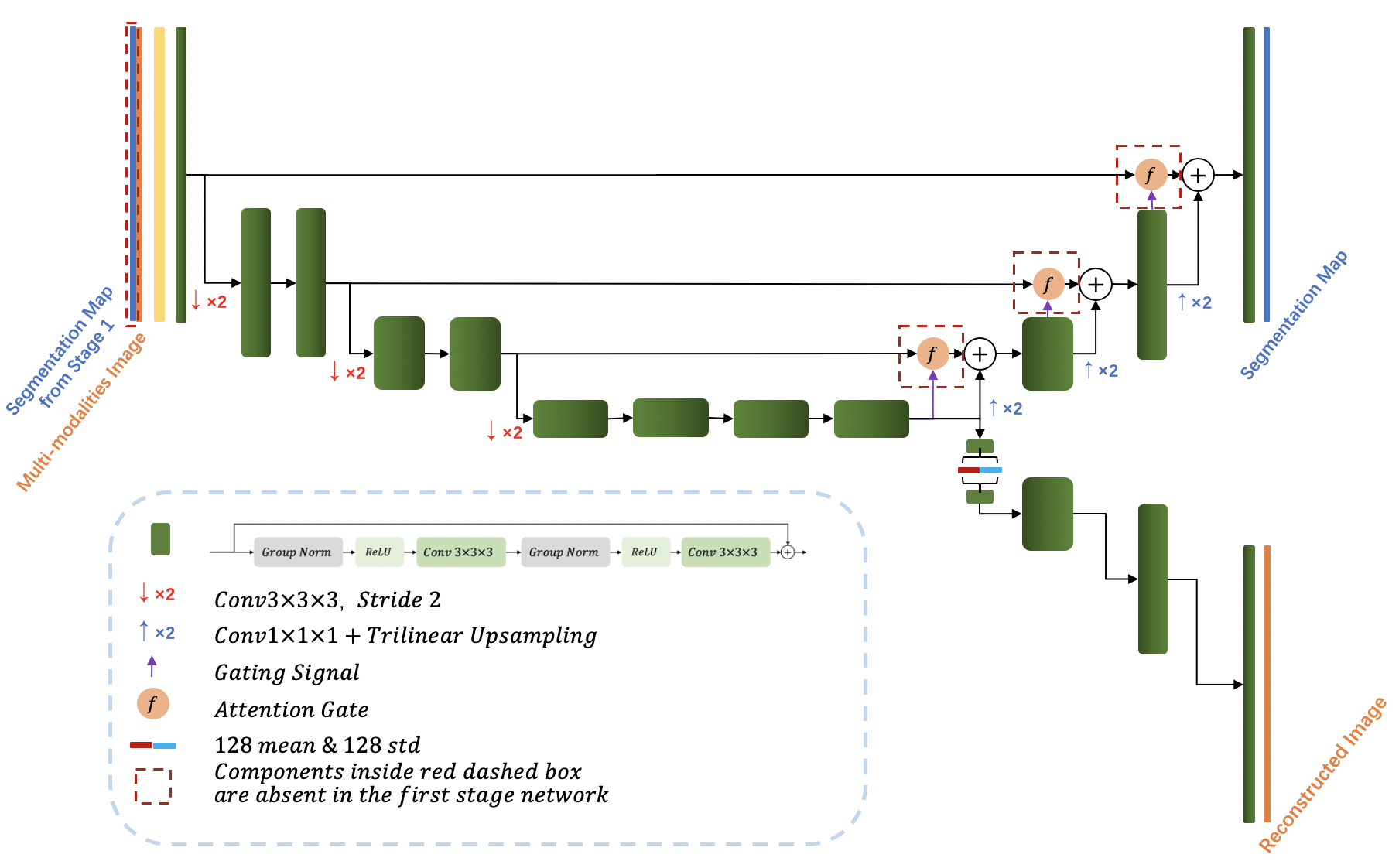}
	\caption{The network architecture of both stages. In the first stage, input (orange strip) is the cropped MRI images ($4\times160\times192\times128$), followed by a $3\times3\times3$ convolution with 32 filters and a dropout layer (yellow strip). The output of the decoder is a segmentation map of size  $3\times160\times192\times128$ with three channels indicating three tumor subregions (WT, TC, and ET). The VAE branch is in charge of input image reconstruction and is disabled while doing inference. In the second stage, the input is the cropped concatenation of the first-stage segmentation map (blue strip) and MRI images (orange strip) (total $7\times128\times128\times128$), and the output is a segmentation map ($3\times128\times128\times128$). Note that there is no input concatenation and attention gates in the first-stage network.}
	\label{fig: Stage1&2}
\end{figure}
\vskip -1.2cm

\subsection{The Second-Stage Network: Attention-Gated Asymmetrical U-Net with the VAE Branch}
\label{sect:Stage2}
The input of the second-stage network (Fig.~\ref{fig: Stage1&2}) is constructed based on the segmentation maps produced by the first-stage network. To alleviate the label imbalance problem, we crop the output of the first-stage network into a spatial size of $128{\times}128{\times}128$ voxels concentrating on the tumor area. The cropped segmentation maps are then concatenated to the original MRI images (cropped to the same area).

\subsubsection{Encoder} The encoder part of the second-stage network has the same structure as in the first-stage network, whereas the input has 7 channels (3 for segmentation maps and 4 for multimodal MRI images), and has a spatial size of $128{\times}128{\times}128$ voxels.

\subsubsection{Decoder} Different from the first-stage network, we add the AGs of \cite{oktay2018attention} in the decoder part. The architecture of the AGs is demonstrated in the next sub-section. At each spatial level, the gating signal from the coarser scale is passed into the attention gate to determine the attention coefficients. The output of an AG is the Hadamard product of input features from encoder through skip connection and attention coefficients. The output of AG at each spatial level is then integrated with the 2-times up-sampled features from the coarser scale by an element-wise summation.  The rest of the network architecture remains the same as the decoder in the first-stage network.

\subsubsection{Attention Gate} 
\begin{figure}[h!]
	\centering
	\includegraphics[width=1\textwidth]{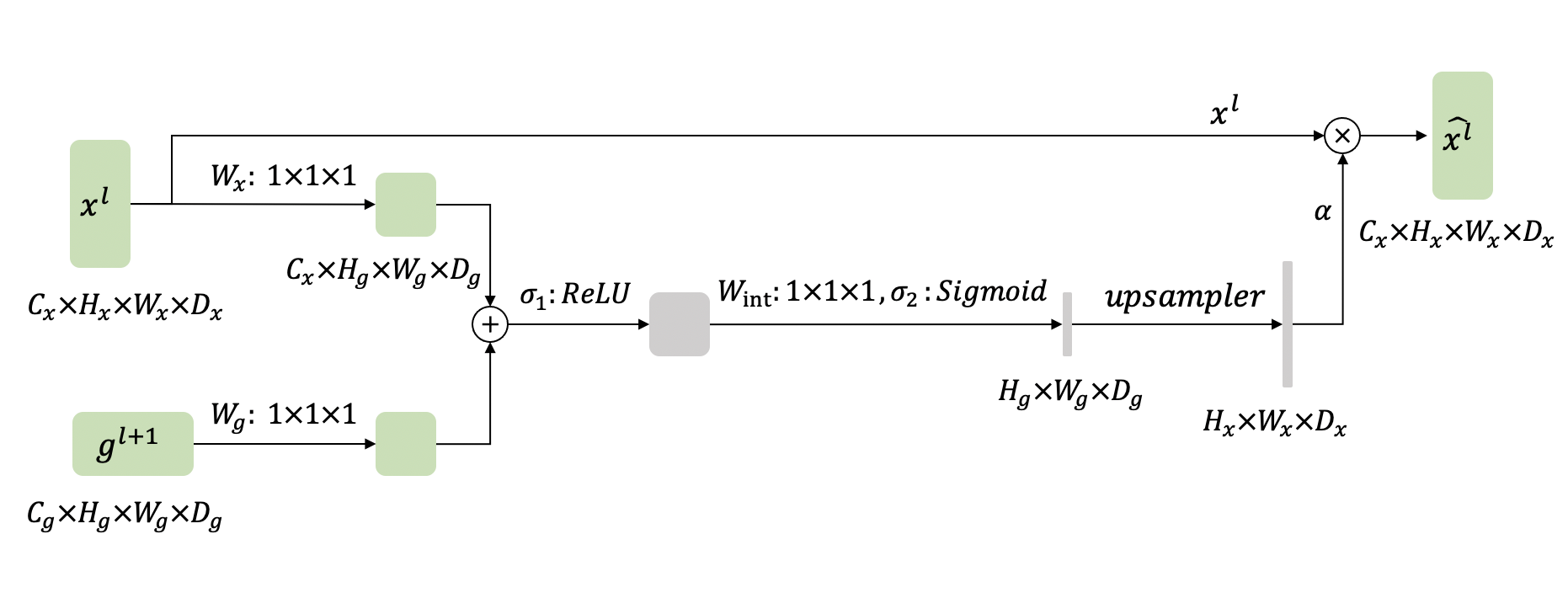}
	\caption{Attention gate.}
	\label{fig: AG}
\end{figure}

Instead of using a single identical scalar value to represent attention level for each pixel vector, a gating vector $g_i$  is computed to determine focus regions for each pixel $i$. Within the $l$-th spatial level, the AG is formulated as follows:
\begin{equation} q^l_{att} = W_{int}^T\sigma_1(W^T_X x^l_i + W^T_g g^{l+1}_i + b_g ) + b_{W_{int}}   \label{ag1}\end{equation}
\begin{equation} \alpha^l_i = \sigma_2(q^l_{att}(x^l_i, g^{l+1}_i;  \theta_{att}))    \label{ag2}\end{equation}
\begin{equation} \hat {x}^l_i =  \alpha^l_i \times x^l_i  \label{ag3}\end{equation}
In each AG (Fig.~\ref{fig: AG}), complementary information is extracted from the gating signal $g^{l+1}_i$ from the coarser scale. To reduce the computational cost, linear transformations $W^T_x$ and $W^T_g$ ($1{\times}1{\times}1$ convolutions) are performed on the input features $x^l_i$ and gating signals $g^{l+1}_i$, to downsize the feature size by 2, and to reduce the number of channels by 2, respectively. The transformed input features and gating signals therefore have the same spatial shape. The sum of them through element-wise summation is activated by the ReLU function $\sigma_1$ and mapped by $W_{int}^T$ into a lower dimensional space for gating operation, followed by the sigmoid function $\sigma_2$ and a trilinear up-sampler to restore the size of attention coefficients matrix $\alpha^l_i$ to match the resolution of the input features. The output $\hat {x}^l_i$ of the AG is obtained by element-wise multiplication of the input features $x^l_i$ and the attention coefficient matrix $\alpha^l_i$. 

\vskip 0.2cm

\subsection{Loss Function}
For both stages, the loss function has 3 parts:
\begin{equation}
L = L_{dice} + 0.1 \times L_{L2} + 0.1 \times L_{KL}. \label{loss}
\end{equation}
$L_{dice}$ is the soft dice loss that encourages the decoder output $p_{pred}$ to match the ground-truth segmentation mask $p_{true}$:
\begin{equation}
L_{dice} = 1 - \frac{2 \times \sum{p_{pred} \times p_{true}}} {\sum{p_{pred}^2} + \sum{p_{true}^2}}. \label{dice}
\end{equation}
$L_{L2}$  is the L2 loss that is applied to the VAE branch output $I_{pred}$ to match the input image $I_{input}$:
\begin{equation}
L_{L2} =  \sum{(I_{pred} - I_{input})^2}.  \label{L2}
\end{equation}
$L_{KL}$ is the KL divergence that is used as a VAE penalty term to induce the estimated Gaussian distribution to approach the standard Gaussian distribution:
\begin{equation}
L_{KL} = \frac{1}{N} \sum {\mu^2} + \sigma^2 - \log{\sigma^2} - 1 , \label{KL}
\end{equation}
where $N$ is the number of the voxels.  As suggested in \cite{myronenko20183d}, we set the hyper-parameter weight to be 0.1 to reach a good balance between the dice and VAE loss terms.

\section{Expriment}\label{Sec: Experiment}
\subsection{Data Description}
The BraTS 2020 training dataset includes 259 cases of HGG and 110 cases of LGG.  All image modalities (T1, T1c, T2, and T2-FLAIR) are co-registered with image size of $240 {\times} 240 {\times} 155$ voxels and 1 mm isotropic resolution.  The training data are provided with annotations, while the validation dataset (125 cases) and testing dataset (166 cases) are provided without annotations. Participants can evaluate their methods by uploading predicted segmentation volumes to the organizer's server. Multiple times of submission for the validation evaluation are permitted, whereas only one submission is allowed for the final testing evaluation.

\subsection{Implementation Details}
Our network is implemented in Pytorch and trained on four NVIDIA P40 GPUs.  
\subsubsection{Optimization} 
We use Adam optimizer with initial learning rate of $lr_0=10^{-4}$ for weights updating. We progressively decay the learning rate according to the following formula: 
\begin{equation}
 lr = lr_0 \times (1 - \frac{e}{N_e})^{0.9},
  \end{equation}
where $e$ is an epoch counter, and $N_e$ is the total number of the epochs during training. In our case, $N_e$ is set to 300.

\subsubsection{Data Preprocessing} Before feeding input data into the first-stage network, we preprocessed the input data by applying intensity normalization to each MRI modality for each patient. The data is subtracted by the mean and divided by the standard deviation of the non-zero region. In the second stage, we crop the segmentation maps from the first-stage network into $128{\times}128{\times}128$-sized patches for each patient while ensuring that the patch includes most tumor voxels. The patches are concatenated with the normalized MRI images (after data augmentation, cropped at the same position) and fed to the second-stage network for training.

\subsubsection{Data Augmentation} To reduce the risk of overfitting, three data augmentation strategies are used. First, the training data is randomly cropped into size of $160{\times}192{\times}128$ before fed into the first-stage network. In addition, in both stages, we randomly shift the intensity of the input data by a value in $[-0.1, 0.1]$ of the standard deviation of each channel, and randomly scale intensity of the input data by a factor in $[0.9, 1.1]$. Finally, we apply random flipping along each 3D axis with a probability of 50\%, in both stages.

\begin{figure}[b!]
	\centering
	\includegraphics[width=1\textwidth]{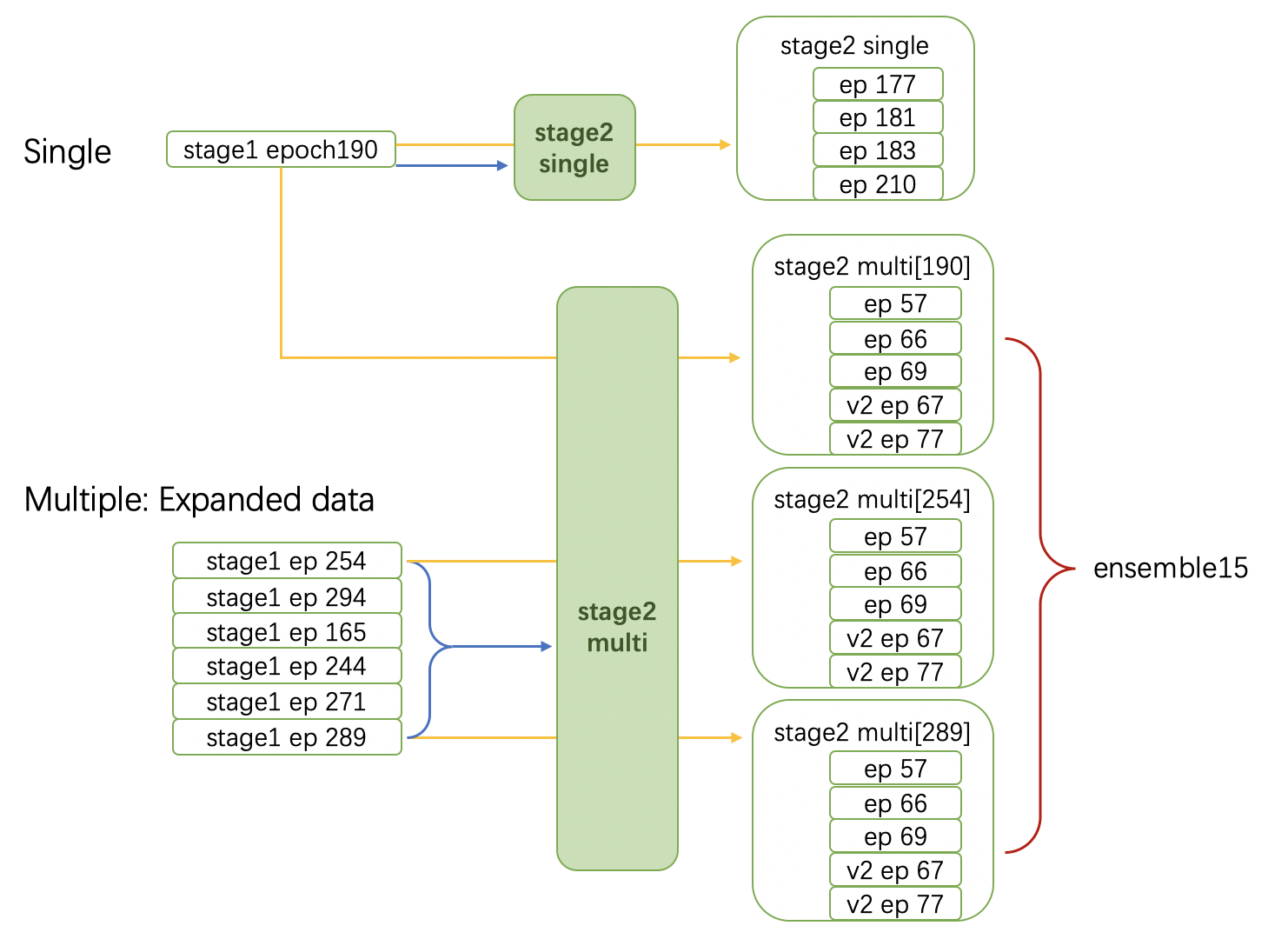}
	\caption{Stage-2 network training schema. In the first case, the training (blue) and inference (yellow) of the stage-2 network are based on the segmentation results of the same stage-1 model; in the second case, the segmentation results from six stage-1 models are combined into an extensive training dataset for training the stage-2 network (blue). The segmentation results of three stage-1 models are respectively input into five stage-2 models of different epochs for inference (yellow), and finally 15 segmentation results are obtained for ensemble (red). }
	\label{fig: schema}
\end{figure}

\subsubsection{Expanded Training Data}
Since the training processes of the two stages are independent, we can select several  first-stage trained models of competitive performance and use their segmentation results as the training data for training the second-stage network. Such a strategy trades a longer training process for better model performance and stability of results. Specifically, we select 6 individual first-stage models (of different epochs with different train-validation divisions) and combined their segmentation results into an extensive dataset to train the second-stage network (Fig.~\ref{fig: schema}). Note that the train-validation division is based on patient IDs. The 6 segmentation results belonging to the same patient are consequentially grouped into the same set. We also have tried training the second-stage network using one single model's segmentation result, but obtained only slight improvement compared to the first-stage network.

\subsubsection{Postprocess}
It is observed that when the predicted volume of ET is particularly small, the algorithm tends to predict TC voxels as ET falsely. In post-processing, based on our experience we replace  ET with TC when the volume of predicted ET is less than 500 voxels. 

\subsubsection{Ensemble }
We use majority voting to conduct model ensembling. In particular, if a voxel has equal votes in multiple categories, the final predicted category of the voxel is determined based on the average probability of each category.

\section{Results}\label{Sec: results}
\subsection{Quantitative Results}
The validation dataset for BraTS 2020 includes 125 cases without providing tumor subtypes (HGG/LGG) or tumor subregion annotations.  Table~\ref{val_result} reports the segmentation result of per-class Dice score and Hausdorff distance for the validation dataset evaluated by the official platform (https://ipp.cbica.upenn.edu/).

\begin{table} [b!]
	\begin{center}
		\caption{Segmentation results on validation data.}\label{val_result}		
		\begin{tabular}{ c<{\centering}|c|c|c|c|c|c|c }
			\hline
			& &\multicolumn{3}{|c|}{\textbf{Dice}} & \multicolumn{3}{|c}{\textbf{Hausdorff (mm)}} \\
			\hline
			\textbf{Stage} & \textbf{Method} & \textbf{ET} & \textbf{WT} & \textbf{TC} & \textbf{ET} & \textbf{WT} & \textbf{TC} \\ \hline
			\multirow{4}*{1} & Model ep190 & 0.7881 & 0.8992 &0.8206 & 23.716 & 5.657& 6.664\\
			& Model ep254 & 0.7930 & 0.8980 & 0.8258 & 26.516 & 5.958 & 6.565\\
			& Model ep289 & 0.7902 & 0.8973 & 0.8286& 24.146 & 6.174 & 7.042\\
			& Ensemble 9 & 0.7946 & 0.9022 & 0.8282 &  23.651	& 5.176 &	6.307 \\
			\hline
			\multirow{7}*{2} & Model ep190: Single & 0.7925 & 0.9012 & 0.8241 & 23.752 & 5.159 & 6.692\\
			&Model ep190: Multiple & 0.7897 & 0.9016 & 0.8291 & 29.327 & 5.288 & 6.632\\
			&Model ep254: Multiple & 0.7769 & 0.9010 & 0.8316 & 32.505 & 5.558&6.557\\
			&Model ep289: Multiple & 0.7896 & 0.9002 & 0.8361 &21.383 & 5.521 & 6.459\\
			& Ensemble 15: Multiple& 0.7960 & 0.9039 & 0.8345 & 23.630 & 4.959 & 6.331 \\
			& Ensemble 21: Multiple& 0.7958 & 0.9041 & 0.8350 & 23.608 & 4.953 & 6.299 \\
			& Ensemble 27: Multiple & 0.7952 & 0.9039 & 0.8350 & 23.590 & 4.962 & 6.303 \\
			\hline
		\end{tabular}
	\end{center}
\end{table}

By comparing the segmentation performance of the $190$th-epoch models of the two stages, we see that the improvement on accuracy  brought by the presence of the second-stage network is more evident for TC than that for WT, and training the second-stage network with expanded training data further improves the Dice score for TC. 

As a performance-boosting component, the second-stage network trained with expanded data can be added to any first-stage model to enhance the segmentation performance. The second-stage network with expanded data also reduces the performance variation across models of different epochs. Table~\ref{var} shows that the standard deviation (SD) of the TC's Dice score and Hausdorff distance are reduced by 68\% and 93\% in the second-stage, respectively. The SDs are calculated based on the performance of all trained non-ensembled models. We also observe that the second-stage network remarkably reduces the variation of ET's Dice score and Hausdorff distance, but this improvement no longer exists after post-processing.

The BraTS 2020 testing dataset contains 166 cases without providing tumor annotations. Our segmentation results on this dataset are presented in Table~\ref{test_result}.

\begin{table} [!]
	\label{table: val_var}
	\begin{center}
		\caption{The performance variation across non-ensembled models on validation data.}\label{var}
	\scalebox{0.83}{		
		\begin{tabular}{ c<{\centering}|c|c|c|c|c|c|c }
			\hline
			& &\multicolumn{3}{|c|}{\textbf{Dice}} & \multicolumn{3}{|c}{\textbf{Hausdorff (mm)}} \\
			\hline
			\textbf{Stage} & \textbf{Metric} & \textbf{ET} & \textbf{WT} & \textbf{TC} & \textbf{ET} &\textbf{WT} & \textbf{TC} \\ \hline
			\multirow{2}*{1} & SD  & 0.0128  & 0.0013  & 0.0110  & 4.097 & 0.236 & 1.427 \\
			& Range & [0.714, 0.750 ]  & [0.894, 0.899]  &  [0.798, 0.835] & [30.535, 42.452] & [5.657, 6.433]& [6.566, 10.502] \\

			\hline
			\multirow{2}*{2} & SD  & 0.0070  & 0.0012 & 0.0035 & 2.337 & 0.184 & 0.102 \\  
			&Range& [0.715, 0.742] & [0.898, 0.902]  &  [0.822, 0.838] & [33.457, 42.546] &  [5.191, 6.024] &[6.168, 6.741] \\  
			\hline
		\end{tabular}
	}
	\end{center}
{\raggedright Note: 
The variation metrics are calculated based on the results from 9 stage-1 models and 37 stage-2 models without post-processing. \par}
\end{table}

\begin{table} [!]
	\label{table: test_result}
	\begin{center}
		\caption{Segmentation results on testing data.}\label{test_result}
		
			\begin{tabular}{ c<{\centering}|c|c|c|c|c|c|c }
			\hline
			& &\multicolumn{3}{|c|}{\textbf{Dice}} & \multicolumn{3}{|c}{\textbf{Hausdorff(mm)}} \\
			\hline
			\textbf{Stage} & \textbf{Method} & \textbf{ET} & \textbf{WT} & \textbf{TC} & \textbf{ET} & \textbf{WT} & \textbf{TC} \\ \hline
			\multirow{1}*{2} 
			& Ensemble 21:Multiple& 0.8205& 0.8729&0.8357 & 15.6711& 11.4288&19.9690 \\
			\hline 
		\end{tabular}
	\end{center}
\end{table}

\subsection{Attention Map}
\begin{figure}[h!]
	\centering
	\includegraphics[width=1\textwidth]{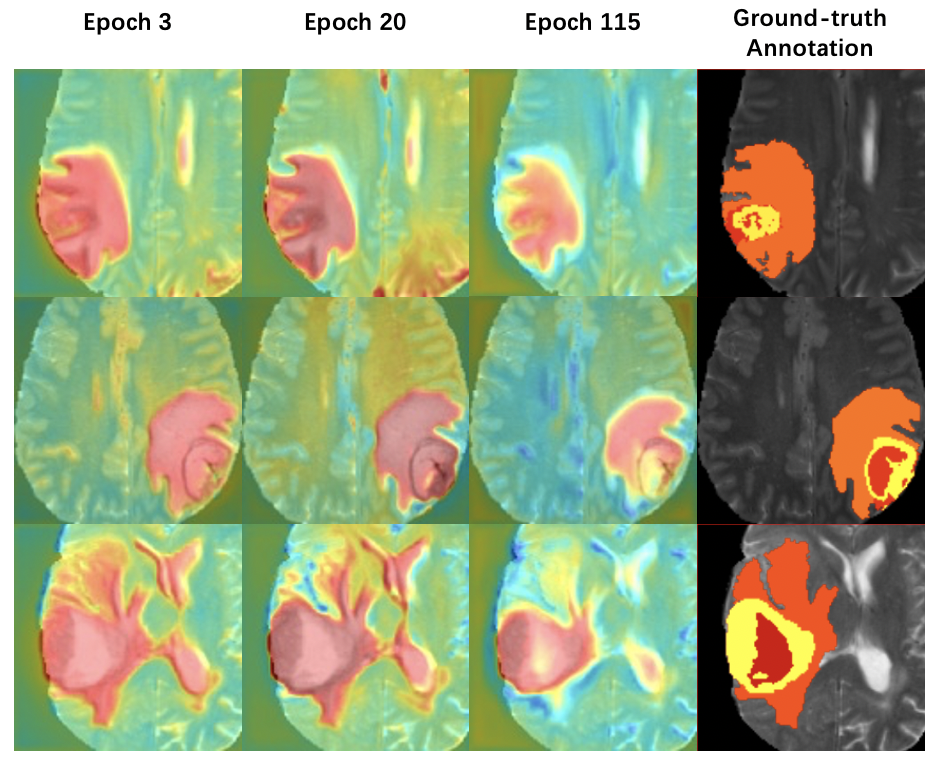}
	\caption{The first three columns show the attention maps at training epochs 3, 20, and 115, respectively. The fourth column shows the example images of T2-modality with ground-truth annotations extracted from the BraTS 2020 training dataset. The model gradually learns to assign lower weights to non-tumor areas and the tumor boundary.}
	\label{fig: Attention}
\end{figure}

The attention matrices in the finest scale are visualised in the form of heatmap with red indicating higher weights and blue indicating lower weights (Fig.~\ref{fig: Attention}). In the first few training epochs, we observe that AGs grasp the tumor's location and meanwhile assign a high weight to gray matter. As the training progresses, the weights assigned to non-tumor regions gradually decrease. AGs also suggest the model avoid misclassification of voxels around the tumor boundary by gradually decreasing weights assigned to the tumor boundary.

\section{Concluding Remarks}
This paper proposes a two-stage cascade network with VAEs and AGs for 3D MRI brain tumor segmentation. The results indicate the second-stage network improves and stabilizes the prediction for all three tumor subregions, particularly for TC and ET (before post-processing).   The second-stage network can also produce more qualified model candidates for further model ensembling. In this study, we use the segmentation results of multiple first-stage models to train the second-stage network. Though this helps improve the model's prediction performance, it noticeably increases the training time as a trade-off. Consequentially, this technique may not be suitable for occasions with limited computing resources and research time. In addition, we can see from Table~\ref{val_result}	 that even if the expanded training data does not include the output of the first-stage 190th-epoch model, we can still use the trained second-stage models to obtain a better result than this first-stage model. This indicates that the second-stage network trained by this strategy has generalizability among models of different epochs.

Since first proposed in natural language processing~\cite{vaswani2017attention}, the attention mechanism has been extensively studied and widely used in image segmentation problems. Technically speaking, the attention mechanism in image segmentation tasks can be divided into the spatial attention, such as the AGs used in our method, and the channel attention, e.g., the ``squeeze and excitation" block in~\cite{hu2018squeeze,zhou2018learning}. It was proposed in \cite{fu2019dual} to combine the two kinds of attention in 2D problems, but multiplications between huge matrices involved in the method will likely exceed the computational limits in 3D scenarios. Further research is expected to include the appropriate combination of the two attention mechanisms into the brain tumor segmentation to enhance the segmentation accuracy. Besides, Dai et al. \cite{dai2018automatic} utilized the extreme gradient boosting (XGboost) in model ensemble and gained extra improvement on accuracy as compared with the majority voting and probability averaging approaches. It may be worth integrating XGboost into our method, as the existence of the second-stage provides more models to be chosen from for the XGboost training.
Moreover, Zhong et al. \cite{zhong20202wm} has recently 
developed a segmentation network model that incorporates the dilated convolution \cite{yu2015multi} and the dense block \cite{huang2017densely}. 
The two popular deep-learning techniques may be valuable
to be combined into our network structure.

\section*{Acknowledgements}
This research was partially supported by the grant R21AG070303
from the National Institutes of Health and a startup fund from 
New York University. 
The content is solely the responsibility of the authors and does not necessarily represent the official views of the National Institutes of Health
or New York University.

\bibliographystyle{splncs04.bst}
\bibliography{report}

\end{document}